\documentclass[proceedings]{JHEP}
\conference{Quantum Aspects of Gauge Theories, Supersymmetry and Unification}
\title{Black Holes and U-duality in Diverse Dimensions}
\author{Mar\'{\i}a A. Lled\'o\\Dipartimento di Fisica, Politecnico di
Torino, \\Corso Duca degli Abruzzi 24, \\10129 Torino, Italy. \\INFN, Sezione
di Torino.}
\abstract{In this paper we review some properties of BPS black holes of
supergravities with $n=32,16$ supersymmetries. The BPS condition, a
condition on the eigenvalues of the central charge matrix, can be shown
to be $U$-duality invariant. We  explicitely work out $D=4,\, N=8$ and
$D=5,\, N=4$ supergravities. }
\begin{document}
\section{Preliminaries}

The purpose of this work is to investigate some properties of black hole
solutions of different supergravity theories in terms of the
corresponding U-duality group.

Black holes are solutions of supergravity theories with point-like
sources. They are rotationally invariant and are, in general, charged
with respect to the vector fields of the theory. We will consider
extremal black hole solutions, that is, solutions on which some of the
supersymmetry charges $Q$ are null. In this case, the solutions are
parametrized by the set of charges $q^\Lambda$ and the values of the
asymptotic values of the scalar fields $\phi^i$ of the theory at
$\infty$ (moduli).

We consider the $N$-extended  super Poincar\'e algebra in a  
$D=4,\dots 9$ dimensional space-time (we follow the notation of Ref.\cite{st}).
 For $D=2n+1$   the irreducible
  representation of
the Clifford algebra $\mathcal{C}(1, D-1)$, of dimension $2^n$,
 is also irreducible with respect to so(1,$D-1$). This is the only
spinor representation and consequently it is self-conjugated. For $D=9$
 it is real, and the anticommutator of two supercharges is given by
 \begin{equation}
\{Q_{\alpha i},Q_{\beta j}\}= \bigl( \Gamma_\mu C p^\mu\bigr)_{\alpha\beta}
\delta_{ij} +C_{\alpha\beta}Z_{ij}
\label{susy9}
\end{equation}
where $i,j=1,\dots N$,  $\alpha,\beta= 1,\dots 2^4$. $C$ is the charge
conjugation matrix that in this case is symmetric, so the central charge
extension of the super Poincar\'e algebra, $Z_{ij}$, is also symmetric.

For $D=5,7$ the spinor representation is pseudoreal, so the
anticommutator is instead
\begin{equation}
\{Q_{\alpha i},Q_{\beta j}\}= \bigl( \Gamma_\mu C p^\mu\bigr)_{\alpha\beta}
\Omega_{ij} +C_{\alpha\beta}Z_{ij},
\label{susy57}
\end{equation}
where $\Omega$ is the symplectic bilinear form. For $D=5$, $C=-C^T$, so
$Z$ is antisymmetric and for $D=7$ $C=C^T$ so $Z$ is symmetric.

For $D=2n$ the
irreducible representation of the Clifford algebra $\mathcal{C}(1,D-1)$
splits into two irreducible pieces under so(1,$D-1$), ({\bf
2}$^{n-1})_\pm$. The projectors on each piece are
$$
\mathcal{P}_\pm= \frac{1\pm \Gamma_{D+1}}{2},\qquad Q_\pm= \mathcal{P}_\pm Q.
$$
For $D=4,8$ these representations are complex and pairwise conjugate.
The anticommutator is
\begin{eqnarray}
\{Q_{\alpha i+},Q^{\dot\beta}_{ j-}\}&=&\bigl(\mathcal{P}_+ \Gamma_\mu C
p^\mu\bigr)^{\dot\beta}_{\alpha}\delta_{ij}\nonumber\\
\{Q_{\alpha i+},Q_{\beta j+}\}&=& (\mathcal{P}_+C)_{\alpha\beta}Z_{ij}.
\label{susy48}
\end{eqnarray}
When $D=4$, $C=-C^T$, so $Z$ is antisymmetric and when $D=8$, $C=C^T$ and
$Z$ is symmetric.

Finally, for $D=6$ the two spinor representations are pseudoreal and
independent. This means that one can choose independently the number of
supersymmetry charges with chirality $+$ or $-$ ($N_+$ and $N_-$).
\begin{eqnarray}
\{Q_{\alpha i+},Q_{ \beta j+}\}&=&\bigl(\mathcal{P}_+ \Gamma_\mu C
p^\mu\bigr)_{\alpha\beta}\Omega_{ij}\nonumber\\
\{Q_{\alpha i+},Q_{\beta j-}\}&=& (\mathcal{P}_+CZ_{ij})_{\alpha\beta}\nonumber\\
\{Q_{\alpha i-},Q_{ \beta j-}\}&=&\bigl(\mathcal{P}_- \Gamma_\mu C
p^\mu\bigr)_{\alpha\beta}\Omega_{ij}.
\label{susy6}
\end{eqnarray}

There are some rotations of the charge vectors  $Q_{\alpha i}$ that
leave invariant the momentum term in the anticommutator of supercharges. These
transformations are automorphisms of the super Poincar\'e algebra, and
they form
the $R$-symmetry group of the algebra.
It is a compact group that we will denote by $H$. The nature of this
group obviously depends on
the reality properties the spinors.  For $D=4,\dots 9$  these are

\begin{center}
{\bf $R$-symmetry group $H$}
\begin{eqnarray*}
&D=9  & \mbox{SO(N)}\\
&D=8  & \mbox{SU(N)}\times \mbox{U(1)}\\
&D=7  & \mbox{USp(N)}\\
&D=6  & \mbox{USp}(N_+)\times\mbox{USp}(N_-)\\
&D=5  & \mbox{USp(N)}\\
&D=4  & \mbox{SU(N)}\times \mbox{U(1)}
\end{eqnarray*}
\end{center}

We consider theories with a maximal number of supesymmetries, $n=32$. 
From the symmetry properties of $Z$ stated above, it follows that the central charge
transforms in the following representation of the $R$-symmetry group,
\begin{center}
{\bf Central charge representation of the $R$-symmetry}
\begin{eqnarray}
&D=9  &\quad \mbox{{\bf 3} of SO(2), ( real symmetric tensor).}\nonumber\\
&D=8  &\quad \mbox{{\bf 3(+)}  of SU(2)$\times$U(1), (complex triplet).}\nonumber\\
&D=7  &\quad \mbox{{\bf 10} of USp(4), (real symmetric tensor).}\nonumber\\
&D=6  &\quad \mbox{{\bf 16} of USp(4)$\times$USp(4)}, \nonumber\\
&& \mbox{(bispinor (4,4) of O(5)$\times $O(5)).}\nonumber\\
&D=5  &\quad \mbox{{\bf 27} of USp(8), ($\Omega$-traceless}\nonumber\\
&&  \mbox{symplectic antisymmetric tensor).}\nonumber\\
&D=4  &\quad \mbox{{\bf 28}  of SU(8), (complex antisymmetric}\nonumber\\
&& \mbox{tensor).}
\label{reph}
\end{eqnarray}
\end{center}

We will see that it is always possible to put the central charge matrix
$Z_{ij}$ in normal form (diagonal or skew diagonal) using a
transformation of the $R$-symmetry group (we denote the eignevalues by $Z_A$).
 Then, using the theory of induced
representations, one can go to the rest frame $(p=(m,0,\dots ,0)$ and the supercharge
anticommutator becomes
\begin{eqnarray*}
\{S_A,S_B^\dagger\}\propto (m+Z_A)\delta_{AB},\\
\{\hat S_A,\hat S_B^\dagger\}\propto (m-Z_A)\delta_{AB},
\end{eqnarray*}
so we obtain a Clifford algebra. It is clear that if $m=|Z_A|$ for some
$A$ (in general one has  $m\geq |Z_A|$), there is one pair of oscillators in the Clifford algebra that
decouple. Representations with such value of the central charge are
representations in which certain supersymmetry charges become trivial . They are called BPS states. If all eigenvalues are equal,
half of the oscillators decouple and we say that we have a
representation with $\frac{1}{2}$BPS. For less restrictive conditions
one obtains representations with  $\frac{1}{4}$BPS or $\frac{1}{8}$BPS.

  \bigskip

The corresponding supergravity theories can be obtained by compactifying
$D=11$, $N=1$ supergravity on a torus $T^{d+1}$ or compactifying $D=10$,
type IIA and IIB supergravity on a torus $T^d$.
The scalar fields of the theory parametrize a coset manifold of the form
$G/H$, where $G$ is a non compact group whose maximally compact subgroup
is $H$. For any $d$, $G=E_{d+1(d+1)}$ \cite{cr}
(the factor U(1) for $D=4$ 
must be supressed to have $H$ as a subgroup of $G$). $G$ is called  the $U$-duality
group. $G$ acts on the non linear manifold of the scalars, and the
vector fields also transform in a representation of $G$. This
representation naturally extends (\ref{reph}) and is given in the
following table for each case \cite{cr},
\begin{center}
   {\bf{ Central charge representation of the  $U$-duality group}}
\begin{eqnarray}
&D=9  &\quad \mbox{{\bf 2+1} of E$_2$=SL(2)$\times$ O(1,1)}\nonumber\\
&D=8  &\quad \mbox{{\bf (3,2)} of E$_3$=Sl(3)$\times$Sl(2)}\nonumber\\
&D=7  &\quad \mbox{{\bf 10} of E$_4$=Sl(5)}\nonumber\\
&D=6  & \quad\mbox{{\bf 16} of E$_5$=O(5,5)}\nonumber\\
&D=5  &\quad \mbox{{\bf 27} of E$_{6(6)}$}\nonumber\\
&D=6  &\quad \mbox{{\bf 56} of E$_{7(7)}$}
\end{eqnarray}\label{udua}
\end{center}

Let $\Phi^i$ be the scalar fields which parametrize $G/H$ and consider
 a local section on the principal bundle $G$ over the
base $G/H$, $L(\phi)\in G$ (coset representative). The action of $G$ can be expressed as
\begin{equation}
L(\phi_g)=gL(\phi)h(\phi_g)
\label{cr}
\end{equation}
where $h(\phi_g)\in H$ is a transformation in the fiber over $\phi_g$.
Choosing a representation of $G$  the coset representative becomes a matrix
$L_a^\Lambda(\Phi)$. The indices $a,\Lambda$ run over the same represenation of $G$
(and of $H$), the different names used to remind the transformation rule
(\ref{cr}).

The charges of a black hole solution can be computed by integrating the
Hodge dual of the field strengths of the vector fields present in the
theory on a $D-2$ spacial
surface enclosing the source. We denote these charges by $q^\Lambda$.
Then, the central charges of the black hole solutions are given by
\begin{equation}
Z_a(q,\Phi)=q^T_\Lambda L^\Lambda_a(\Phi)
\label{cc}
\end{equation}
Notice that the vectors fields  are in a
certain representation of $G$, which necesarily coincides with the
  representation in (\ref{udua}).

When the central charges correspond to BPS states,  some
supersymmetry generators are null when acting on the black hole
solutions, so the solutions are "supersymmetric". The
condition to be a BPS black hole is a condition on $Z$, that is,  on $q$ and $\Phi$. 
  Due to the form of (\ref{cc}), it is clear that any condition 
$E_\alpha(Z)=0$ (where $\alpha$ runs over some representation $T$ of $G$)
 that  is covariant under $G$, that is
\begin{equation}
E_\alpha(Zg)=E_\beta(Z) T(g)^\beta_\alpha,
\label{covc}
\end{equation}
will become simply a condition on $q$, $E_\alpha(q)=0$. It was shown in
\cite{dfl} that this condition is actually moduli
independent or $U$-duality invariant  in all theories in dimensions
$D=4,\dots ,9$ with $n=32,16$ supersymmetries. Here we will review these
results, in particular $D=4$, $N=8$ and $D=5$, $N=4$.

\section{Maximal supergravity in dimensions $D=4,\dots, 9$}

We study the diagonalization of the matrix of  central charges in the
different cases (Table \ref{ccm}).
 \TABULAR{lllll}{
 $D$& $N$ & $H$ & Z&  eigenvalues\\
 9&2&SO(2)& real symmetric.&2\\
 8&2&SU(2)$\times$U(1)& complex symmetric&2\\
 7&4&USp(4)$\approx$O(5)&quaternionic symmetric&2\\
 6&4,4&USp(4)$\times$USp(4)&quaternionic&4\\
 5&8&USp(8)& quaternionic antisymmetric&4\\
 4&8&SU(8)& complex antisymmetric&4 
 }{Central charge matrix\label{ccm}}
 A matrix is quaternionic (or symplectic) if
 $$
 Z^*=-\Omega Z\Omega.
$$
A matrix satisfying this conditions can be understood as a matrix whose
entries are quaternions, and these are in the representation
$$\mbox{Id}_{2\times 2},\quad i\sigma_1, \quad i\sigma_2,\quad i\sigma_3.$$

One has the following results:

\smallskip

\noindent 1. Any matrix can be brought to a diagonal
form by making a transformation
$$Z_D=U_1ZU^\dagger_2$$
where $U_i$ are orthogonal matrices if $Z$ is real, unitary  if $Z$ is
complex and unitary symplectic if $Z$ is quaternionic (In the
representation of $n\times n$ quaternionic matrices by complex,
$2n\times 2n$ matrices, it is $\Omega Z$ which is brought to diagonal form).

\noindent 2. In the case  that the original matrices are symmetric, hermitian or symplectic
hermitian respectively, $U_1=U_2$ and the eigenvalues are real.

\noindent 3. An antisymmetric matrix can be brought to skew diagonal form by a
transformation
$$
Z_{SD}=UZU^T
$$
where $U$ belongs to the appropriate group as before and the eigenvalues
are real.

\smallskip

For $D=9$ we can trivially diagonalize $Z$ with an $R$-symmetry transformation by
using the result 2. For $D=8$, an $R$-transformation $U$ brings the
matrix $Z$ to $UZU^T$.  It is easy to see that to diagonalize the
$2\times 2$ matrix it is enough to use the
result 1 with $U_2^\dagger =U_1^T$, being the eigenvalues real.
 For $D=7$, $Z$ is four
dimensional and since the spinors are pseudoreal it is quaternionic.
Since it is also symmetric, we can take advantage of the isomorphism
Sp(4)$\approx$O(5) and decompose $Z$ as
$$Z_{ij}=Z_{IJ}(\gamma^{IJ})_{ab}, IJ=1,\dots 5
$$
where $Z^{IJ}$ is real and antisymmetric and
$$\gamma^{IJ}=\frac{1}{2}[\gamma^I,\gamma^J],
$$
and $\gamma^I$ are the gamma matrices of O(5). We can skew diagonalize
$Z^{IJ}$ using the  result 3, with only  two
independent eigenvalues.  For $D=6$ one can use the result 1 and
diagonalize the matrix with an element
$(U_1,U_2)\in$USp(4)$\times$USp(4). The quaternionic property is
 preserved by a transformation $Z_D=U_1ZU^\dagger_2$, but since the
eigenvalues are not real there are four
independent real quantities.  Finally, for $D=5, 4$ one can use the result 3.

For $D=7,8,9$ there are only two independent eigenvalues. In the generic
situation, when  the two eigenvalues are different, we have that the
bound $m=|Z_A|$ can be reached only by  the highest eigenvalue. Then,
$\frac{1}{4}$ of the oscillators of the supersymmetry alegra are null
on the solution. We say that we have a $\frac{1}{4}$BPS state. If all
the eigenvalues are equal, the half of the oscillators decouple and we
have a $\frac{1}{2}$BPS state. For $D=4,5,6$ we have four eignevalues
and the generic case, when all of them are different preserves
$\frac{1}{8}$ of the supersymmetry. If the eigenvalues are equal by
pairs we have $\frac{1}{4}$BPS states and if they are all equal we have
$\frac{1}{2}$BPS states. Exotic cases like having only one pair of equal
eigenvalues and the rest different or having three equal eigenvalues and
a different fourth one are forbidden on physical grounds as we will see
in the next example.

\subsection{ BPS states in $D=4$, $N=8$ Supergravity.}

As we have seen in this case $Z$ is a $8\times 8$ complex antisymmetric 
matrix, so it can be skew diagonalized
$$
Z_{SD}=UZU^T
$$
with 4 independent eigenvalues. Since we can use only transformations
$U\in $ SU(8), the eigenvalues are not real. Instead  there is an overall phase 
that cannot be removed \cite{fm}. In fact, this phase is an extra
parameter of the solution which usually  is set to zero and
 all the eigenvalues real. Their absolute value,
 can be computed as the square root of the eigenvalues of the matrix $ZZ^\dagger$.

\paragraph{$\frac{1}{2}$BPS}

Since all the eigenvalues are real, the matrix $ZZ^\dagger$ is a
multiple of the identity,
\begin{equation}
ZZ^\dagger=\mbox{Tr}(ZZ^\dagger)\frac{1}{8}\mbox{Id}.
\label{su8cc}
\end{equation}
This is an SU(8) covariant constraint. The idea is to find an E$_{7,7}$
covariant constraint that necessarily implies (\ref{su8cc}). Then we
will be in the situation (\ref{covc}), where the $\frac{1}{2}$BPS condition
is  moduli independent. Nevertheless,  the eigenvalues will 
depend on the moduli.

We consider the irreducible representation of E$_{7,7}$ {\bf 56} (the
number indicates the  dimensionality of the representation). Under SU(8) it decomposes  as
${\bf 56=28 +\bar{28}}$. ${\bf 28}$ is the twofold antisymmetric
representation of SU(8) in which $Z$ sits.  We will write a vector of {\bf
56} symbolically as $\tilde Z=(Z,\bar Z)$. We consider now the quartic
invariant of this representation of E$_{7,7}$ \cite{fg, adf,kk},
\begin{equation}
I=4\mbox{Tr}(Z\bar Z)^2 -(\mbox{Tr}Z\bar Z)^2+2^4({\it Pf}Z+{\it Pf}\bar Z),
 \label{quartic}
\end{equation}
where
$$
{\it Pf}Z={1\over 2^44!}\epsilon^{ABCDRPGH}Z_{AB}Z_{CD}Z_{RP}Z_{GH}
$$
is the Pfaffian of $Z$. Let us take the second derivative of this invariant
$$
{\partial^2 I\over \partial \tilde Z\partial \tilde Z}.
$$
It is a quadratic polynomial which is a symmetric tensor, so it sits  in the
$({\bf 56}\times {\bf 56})_S={\bf 1596}$
representation of E$_7$ which is reducible and decomposes as {\bf 1463
+ 133}. {\bf 133} is the adjoint of E$_7$, so we can take the projection 
 \begin{equation}
 {\partial^2 I\over \partial \tilde Z\partial \tilde Z}|_{Adj_{E_7}}.
 \label{proj}
\end{equation}
Since {\bf  133} decomposes as {\bf 63+70} under SU(8) ({\bf 63} is the
adjoint representation of SU(8)), the expression
(\ref{proj})  splits into  two  SU(8) covariant
polynomials
\begin{eqnarray}
&&V_A^C=\frac{\partial^2 I}{\partial Z_{AB}\bar\partial Z^{CB}}|_{Adj_{SU(8)}}\approx
\nonumber\\
&&( Z_{AB}\bar Z^{CB}-\frac{1}{8}\delta_A^C Z_{PQ}\bar
Z^{PQ}),\label{vac}\\
&&V^+_{[ABCD]}=\frac{\partial^2 I}{ \partial Z_{AB}\bar \partial Z^{CB}}|_{\bf
70}\approx
\frac{\partial^2 I}{\partial Z_{[AB}\partial Z_{CD]}}-\nonumber\\
&&\frac{1}{4!}\epsilon^{ABCDPQRS}
\frac{\partial^2 I}{\partial \bar Z^{[AB}\partial \bar Z^{CD]}}.
\label{selfdual}
\end{eqnarray}
The {\bf 63} piece, (\ref{vac}) is just (\ref{su8cc}), which is the
constraint we want. But the {\bf 70} part $V^+_{[ABCD]}$ is not, in principle,
zero. Here is where the $G/H$ structure of the theory enters. Consider
$L(\Phi):G/H\mapsto G$, the local section or coset representative of
$G/H$. Using this map we can make  a pull-back of the Maurer-Cartan left
invariant forms on $G$ to an open set of $G/H$,
\begin{equation}
\Omega(\phi)=L^{-1}(\phi)dL(\phi)=\omega_iT_i +P_\alpha T_\alpha.
\label{mc}
\end{equation}
$T_i\in \mathcal{H}$ are the  generators in the Lie algebra of $H$ and
$T_\alpha\in \mathcal{K}$, where $\mathcal{G}=\mathcal{H}+\mathcal{K} $
is a Cartan decomposition of  $\mathcal{G}$, the Lie algebra of $G$.
$\omega_i$ are the components of the spin connection of the bundle
$G\mapsto G/H$ and $P_\alpha$ is the vielbein of the invariant metric on $G/H$.

In the case of E$_{7,7}$/SU(8), the equation (\ref{mc}) takes the form
$$
\nabla_{SU(8)}Z_{AB}={1\over 2}\epsilon_{ABCD}\bar Z^{CD}.
$$
Taking the covariant derivative $\nabla_{SU(8)}$  of the equation
$V_A^C=0$ and using the Maurer-Cartan equations above, one obtains
$$
V_+^{ABCD}=0,
$$
as we wanted to show.

Notice that the reality of the eigenvalues follows in fact from
(\ref{selfdual}), so it follows from the E$_{7,7}$ invariance.

\paragraph{ $\frac{1}{4}$BPS.}

In this case we have that the eignevalues are equal by pairs.  It can be
shown \cite{dfl} that the E$_{7,7}$ covariant condition for this to
happen is
$$
 {\partial I\over \partial Z_{AB}}=0\quad(\Rightarrow
 {\partial I\over \partial\bar Z^{AB}}=0),
 $$
where $I$ is again the quartic invariant. The reality also
follows from the E$_{7,7}$ invariance.

\medskip
$\frac{1}{8}$BPS is the generic case when all the eigenvalues are
different. One can always make the highest eigenvalue (which is the mass
of the BPS state) real, so no equation is needed to assure reality.
Having  three or two eigenvalues  equal and the
rest different makes the quartic invariant negative \cite{f}. Since this
invariant has an interpretation as the entropy of the black hole
squared, $I\propto S^2$ it cannot be negative, so these cases are excluded.

For all the other theories in Table \ref{ccm} a similar analysis can be
carried out \cite {dfl, fm, fg, lsp}

\section{Supergravities with 16 supersymmetries. BPS states of $D=5$, $N=4$ supergravity}

Theories with sixteen supersymmetries are obtained  as compactifications of
the heterotic string on tori T$^d$ ($1\leq d\leq 6$). They can
also be obtained by compactifying $D=11$ supergravity or $D=10$   
Type IIA and IIB supergravity on manifolds preserving less
supersymmetries,  as    K$_3$. 
In these theories there are matter fields and the $U$-duality group $G$
depends on the matter content as well as on the dimension of space-time.
The maximal compact subgroup is a direct product $H_R\times H_M$ where
$H_R$ is the R-symmetry of the corresponding supersymmetry algebra
and $H_M$ is the group
acting on the matter multiplets. If $n$ is the number of matter
multiplets this group is $H_M=\mbox{O}(n)$. $G$ is of the form O(10-$D,n$)$\times$O(1,1)
for $5\leq D\leq 9$ while for $D=4$ it is SL(2)$\times$O(6,$n$). The R-symmetry  groups are
O(10-$D$) for $5\leq D\leq 9$ and O(6)$\times$O(2)$\approx$SU(4)$\times$U(1) for $D$=4.

The $G$ and $H_R$ representations of the central charges are given in the following tables.

\begin{center}
{\bf Central charge representation of $H_R$.}
\begin{eqnarray*}
D=9 &\quad {\bf 1}\quad  & \mbox{O}(1)=\mbox{Id}\\
D=8 &\quad {\bf 1^c}\; \mbox{complex} & \mbox{U}(1)\approx \mbox{O}(2)\\
D=7 &\quad {\bf 3}\; \mbox{real} & \mbox{SU}(2)\approx \mbox{USp}(2)\\
D=6 &\quad {\bf 4}\; \mbox{real} &  \mbox{O}(4)\approx \mbox{USp}(2)\times\mbox{USp}(2)\\
D=5 &\quad {\bf 1+5}\; \mbox{real}&  \mbox{O}(5)\approx \mbox{USp}(4)\\
D=4 &\quad {\bf 6^c}\; \mbox{complex}  & \mbox{O}(6)\times\mbox{O}(2)\approx \\
&&\mbox{SU}(4)\times\mbox{U}(1)
\end{eqnarray*}
\end{center}

From the above table, and according  to our previous analysis,
it follows that the matrix of central charges,  $Z$, has  only one 
independent real eigenvalue
for $D=6,\dots 9$ and two independent eigenvalues for $d=5,6$. Therefore, for $D=6,\dots 9$
only 1/2 BPS states can occur while for $D=4,5$ both, 1/2 and 1/4 BPS states can occur.

\begin{center}
{\bf Central charge representation of $G$}
\begin{eqnarray*}
&D=6,\dots 9  &{\bf d+n}\; \mbox{real vector}\\
& &\mbox{O}(d,n)\times \mbox{O}(1,1)\nonumber\\
&D=5 & {\bf 1+(5\!+\! n)}\; \mbox{(singlet+vector)} r\\
&&  \mbox{O}(5,n)\times\mbox{O}(1,1)\\
&D=4 & {\bf(2, 6\!+\!n) }\;\\
&& \mbox{Sl}(2)\times\mbox{SO}(6,n)
\end{eqnarray*}
\end{center}

We will briefly outline here the case of $D=5$. It corresponds to
heterotic string on $T^5$ or $D=11$ supergravity on $K^3\times T^2$. The number of matter
multiplets is $n=21$, although our analysis is independent of such number.

The central charge $\hat Z$ is  an antisymmetric quaternionic matrix.
This implies that the matrix $Z=\hat Z\Omega$ is hermitian and
quatenionic. The  $4\times 4$ matrix depends on 6 real parameters. We
want to
exploit the isomorphism O(5)$\approx$USp(4), so we decompose $Z$ as
$$
Z=Z^a\gamma_a +Z^0\mbox{Id},
$$
where $\gamma_a$, $a=1,\dots 5$ are the O(5) $\gamma$-matrices and $Z^a, Z^0$ are
real numbers. $Z^0$ is a singlet under O(5) and $Z^a$ is a vector.
It follows that
\begin{eqnarray}
&&\mbox{Tr}Z=4 Z^0\nonumber \\
&&(\mbox{detZ)}^{1/2}=  {1\over 8}(\mbox{Tr}Z)^2-{1\over 4}\mbox{Tr}Z^2=\nonumber\\
&&{Z^0}^2-\vec{Z}^2
\label{trdet}\end{eqnarray}
The  characteristic equation of $Z$ (or better, its square root) is
\begin{eqnarray*}
&&\sqrt{\mbox{det}Z-\lambda\mbox{Id}}=\\
&&\lambda^2-{1\over 2}\mbox{Tr}Z\, \lambda +(\mbox{det}Z)^{1/ 2}=0,
\end{eqnarray*}
implying that  $Z$ has two coinciding eigenvalues (in absolute value) either if
$$
\mbox{Tr}Z=0 \quad \mbox{or}\quad {1\over 4}(\mbox{Tr}Z)^2=4(\mbox{det}Z)^{1/2}
$$
Using (\ref{trdet}), the above equation directly implies
\begin{equation}
Z_0Z_a=0. \label{condi}
\end{equation}
The eigenvalues are given by
$$
\lambda_{1,2}={1\over 2}\left({1\over 2}\mbox{Tr}Z\pm \sqrt{\mbox{Tr}Z^2-{1\over 4}
(\mbox{Tr}Z)^2}\right),
$$
being the plus sign the mass squared of the BPS state.

The central charge representation of $G$ (and $H$), singlet + vector, is reducible and
the coset representative $L^\Lambda_a$ splits into blocks
$$
\pmatrix{e^{2\sigma}& \cr
&e^{-\sigma}\pmatrix{M}},
$$
where $\sigma$ parametrizes O(1,1) and $M$ is in the fundamental
representation of O(5,n). Since $Z_a=q_\Lambda L^\Lambda _a$,
then $Z_0=e^{2\sigma}q_0$.  The condition $Z_0=0$ implies $q_0=0$, which is a singlet
of O(5+n), and then, $U$-duality invariant.

If $Z_I,\; I=1,\dots n$ are the matter charges associated to the $n$
matter multiplets, we have that, because of (\ref{mc}),
$$
\nabla_{O(5)}Z_a=\mbox{Tr}(\gamma_aP_I)Z^I +Z_ad\sigma, 
$$
 therefore $Z_a=0$ implies $Z_I=0$.
This is also an O(5,n) invariant statement since, it comes by differentiating the quadratic
invariant polynomial
$$
I=\sum_{a=1}^5Z_aZ^a- \sum_{I=1}^MZ_IZ^I.
$$
Therefore, $Z_a=Z_I=0$ implies $q^\Lambda=0$ where $q^\Lambda$,
$\Lambda=1,\dots 5+n$, is a fixed charge vector of O(5,M),
as found in \cite{fm}.

\medskip

\noindent{\large\bf Acknowledgments}

\medskip

I want to thank to my collaborators R. D'Auria and S. Ferrara
for reading the manuscript an suggesting many improvements.

\end{document}